\documentclass[11pt]{article}\usepackage[]{graphicx}\usepackage[]{color}
\makeatletter
\def\maxwidth{ %
  \ifdim\Gin@nat@width>\linewidth
    \linewidth
  \else
    \Gin@nat@width
  \fi
}
\makeatother

\definecolor{fgcolor}{rgb}{0.345, 0.345, 0.345}

\usepackage{framed}
\makeatletter
 {\par\unskip\endMakeFramed%
 \at@end@of@kframe}
\makeatother

\definecolor{shadecolor}{rgb}{.97, .97, .97}
\definecolor{messagecolor}{rgb}{0, 0, 0}
\definecolor{warningcolor}{rgb}{1, 0, 1}
\definecolor{errorcolor}{rgb}{1, 0, 0}
\newenvironment{knitrout}{}{}

\usepackage{alltt}
\usepackage{natbib,amssymb,amsmath,amsthm,fullpage,hyperref,graphicx}

\numberwithin{equation}{section}
\theoremstyle{plain}

\newcommand{\bl}[1]{{\mathbf #1}}
\newcommand{\bs}[1]{{\boldsymbol #1}}
\newcommand{\Exp}[1]{{\text{E}}[ \ensuremath{ #1 } ]  }
\newcommand{\Var}[1]{{\text{Var}}[ \ensuremath{ #1 } ]  }
\newcommand{\Cov}[1]{{\text{Cov}}[ \ensuremath{ #1 } ]  }

\title{Additive and multiplicative effects network models}
\author{Peter D. Hoff \\
Department of Statistical Science \\
Duke University}
\date{\today}
\IfFileExists{upquote.sty}{\usepackage{upquote}}{}
\begin{document}

\maketitle

\begin{abstract} 
Network datasets typically exhibit 
certain types of statistical dependencies, 
such as 
within-dyad correlation, row and column heterogeneity, 
and third-order dependence patterns such as transitivity 
and clustering. The first two of these  can 
be well-represented statistically with 
a social relations model, a type of additive 
random effects model originally developed for 
continuous dyadic data. Third-order patterns 
can be represented with multiplicative random effects
models, which are related to 
matrix decompositions 
commonly used for matrix-variate data analysis. 
Additionally, these multiplicative random effects models 
generalize other popular latent variable network 
models, such as the stochastic blockmodel and 
the latent space model. 
In this article
we review a general regression framework for the analysis 
of network data that combines these two types 
of random effects
and accommodates a variety of 
network data types, including 
continuous, binary and ordinal network relations. 

\smallskip

\noindent {\it Keywords:}
Bayesian, 
factor model,
generalized linear model,
latent variable,
matrix decomposition,
mixed effects model. 
\end{abstract}

\section{Introduction}
Network data provide quantitative information 
about relationships among objects, individuals or 
entities, which we 
refer to as nodes. 
Most network data  
quantify pairwise relationships between
nodes. A pair of nodes is referred to 
as a \emph{dyad}, and a quantity 
that is measured or observed for multiple dyads 
is called a \emph{dyadic variable}. Common sample spaces for 
dyadic variables include continuous, discrete, dichotomous 
and ordinal spaces, among others. 
Examples of dyadic variables include quantitative measures of
trade flows between countries, communications among people,
binding activity among proteins, and structural connections among
regions of the brain, to name just a few.   

Measurements of a dyadic variable 
on a population 
of $n$ nodes may be summarized with a \emph{sociomatrix}, an $n\times n$ 
square matrix $\bl Y$ with an undefined diagonal, where 
entry $y_{i,j}$ denotes the value of the relationship 
between nodes $i$ and $j$ from the perspective of node $i$, 
or in the direction from $i$ to $j$. 
Analysis of an observed sociomatrix $\bl Y$ 
often proceeds in the context of one or more 
statistical models, with which a data analyst 
may evaluate competing theories of network formation, 
describe patterns in the network, estimate 
effects of other variables on dyadic relations, 
or impute missing values. 

While most of the dyadic variables I have encountered are not dichotomous 
in their raw form, much of the statistical literature has focused 
on binary network data for which the sociomatrix 
$\bl Y$ can be viewed as the adjacency matrix of a graph. 
Many statistical random graph models are motivated by intuitive, 
preconceived notions of how networks may form, 
particularly social networks. 
For example,  preferential attachment models 
view an observed network as the end result of a 
social process in which nodes are sequentially introduced into 
a population of existing nodes \citep{price_1976}. As another example, 
the parameters in 
the types of exponential family graph models that are commonly used
have interpretations as  node-level preferences 
for certain relationship outcomes 
\citep{wasserman_pattison_1996}.

An alternative approach is to build a statistical model 
for $\bl Y$ based on its inherent structure as a sociomatrix, 
that is, as a data matrix whose row labels are the same as its 
column labels. Such an approach  can build upon
familiar, well-developed statistical methodologies 
such as ANOVA, linear regression, matrix decompositions, factor analysis  and 
linear and generalized linear mixed effects models, 
and can be applied to a wide variety of dyadic data types. 
In this article, we review such a framework for network data 
analysis using these tools, starting with simple ANOVA-style decompositions 
of sociomatrices and ending with 
additive and multiplicative random effects regression models  
for continuous, binary, ordinal and other types of 
dyadic network data. 

In the next section we review an ANOVA-style decomposition of 
a sociomatrix known as the social relations model (SRM)
\citep{warner_kenny_stoto_1979,wong_1982}, 
which corresponds to 
a particular
Gaussian additive random effects model for network data. An extension 
of this model that includes  covariates is also developed, 
which we call the social relations 
regression model (SRRM). 
The SRM and SRRM are able to describe network variances and covariances, 
but are unable to describe third-order 
dependence patterns such as 
transitivity, balance, or the existence of clusters of 
nodes with high subgroup densities of ties. 
In Section 3 we discuss how such patterns can be represented 
by a multiplicative latent factor model, in which 
the relationship between two nodes depends on the similarity 
of their unobserved latent factors. 
From a matrix decomposition perspective, this motivates 
the use of an ``additive main effects, multiplicative interaction'' (AMMI)
matrix  model \citep{gollob_1968, bradu_gabriel_1974}. 
Combining an AMMI model 
 with a social relations covariance model yields 
what we call an additive and multiplicative effects (AME) 
 network model. 

These AME models are built from 
linear regression, 
random effects models
and matrix decomposition - methods which are most appropriate
for continuous data consisting of a signal of interest plus 
Gaussian noise. 
In contrast,
many dyadic variables are
discrete, ordinal, binary or sparse. In Section 4 we extend the AME 
framework to accommodate these and other types of dyadic variables 
using a Gaussian 
transformation model. 
In Section 5 we compare the multiplicative effects component 
of an AME model with two other latent variable network models, 
the stochastic blockmodel \citep{nowicki_snijders_2001}
and the latent space model \citep{hoff_raftery_handcock_2002}. 
We review results showing that these latter two models 
can be viewed as submodels of the multiplicative effects 
model.  Connections to exponentially parameterized 
random graph models  (ERGMs) \citep{wasserman_pattison_1996} 
are also discussed. 
Section 6 presents a Markov chain Monte Carlo algorithm 
for Bayesian model fitting of a hierarchy of 
AME network models.  A discussion follows 
in Section 7.

\section{Social Relations Regression}
\subsection{ANOVA and the Social Relations Model}
Numeric sociomatrices typically 
exhibit 
certain statistical features.
For example, it is often  the case that
values of the dyadic variable in a given row of the sociomatrix are 
correlated with one another, in the 
sense that high and low values 
are not equally distributed among the rows, resulting in 
substantial heterogeneity of the row means of the sociomatrix. 
Such heterogeneity can be explained by the fact that 
the relations within a row all share a
common ``sender,'' or row index.
If sender $i_1$ is more ``sociable'' than sender $i_2$, we would
expect the values in row $i_1$ to be larger than
those in row $i_2$, on average.
In this way, heterogeneity of the nodes in terms of their
sociability contributes to an across-row
variance of the row means of the sociomatrix.
Similarly,
nodal heterogeneity in ``popularity'' contributes to
the across-column variance of the column means.

A classical approach to evaluating across-row and across-column
heterogeneity in a data matrix is the ANOVA decomposition.
A statistical model based on the ANOVA decomposition posits that
the variability of
the $y_{i,j}$'s around some overall mean $\mu$ is well-represented by
additive row and column effects: 
\begin{align}
y_{i,j} = \mu + a_i +b_j + \epsilon_{i,j}.
\label{eqn:rce}
\end{align}
In this model,
heterogeneity among 
the $a_i$'s and $b_j$'s 
gives rise to
observed heterogeneity in the row means
and column means of the sociomatrix, respectively.

While straightforward to implement, a classical ANOVA analysis ignores
a fundamental characteristic of dyadic data: Each node
appears in the dataset as both a sender and a receiver of
relations, or equivalently, the
row and column labels of the data matrix refer to the same set of
nodes.
In the context of the ANOVA model,
this means that
each node $i$ has two additive effects: a row effect
$a_i$  and a column effect $b_i$. Since 
each pair of effects $(a_i,b_i)$ 
shares a node, 
a correlation between the vectors 
$(a_1,\ldots, a_n)$ and $(b_1,\ldots, b_n)$ 
may be expected. 
Additionally,
each  dyad $\{i,j\}$ has two
outcomes, $y_{i,j}$ and $y_{j,i}$.
As such, the possibility that 
$\epsilon_{i,j}$ and $\epsilon_{j,i}$ are correlated should be considered.

\begin{figure}
\begin{knitrout}\footnotesize
\definecolor{shadecolor}{rgb}{0.969, 0.969, 0.969}\color{fgcolor}

{\centering \includegraphics[width=6in]{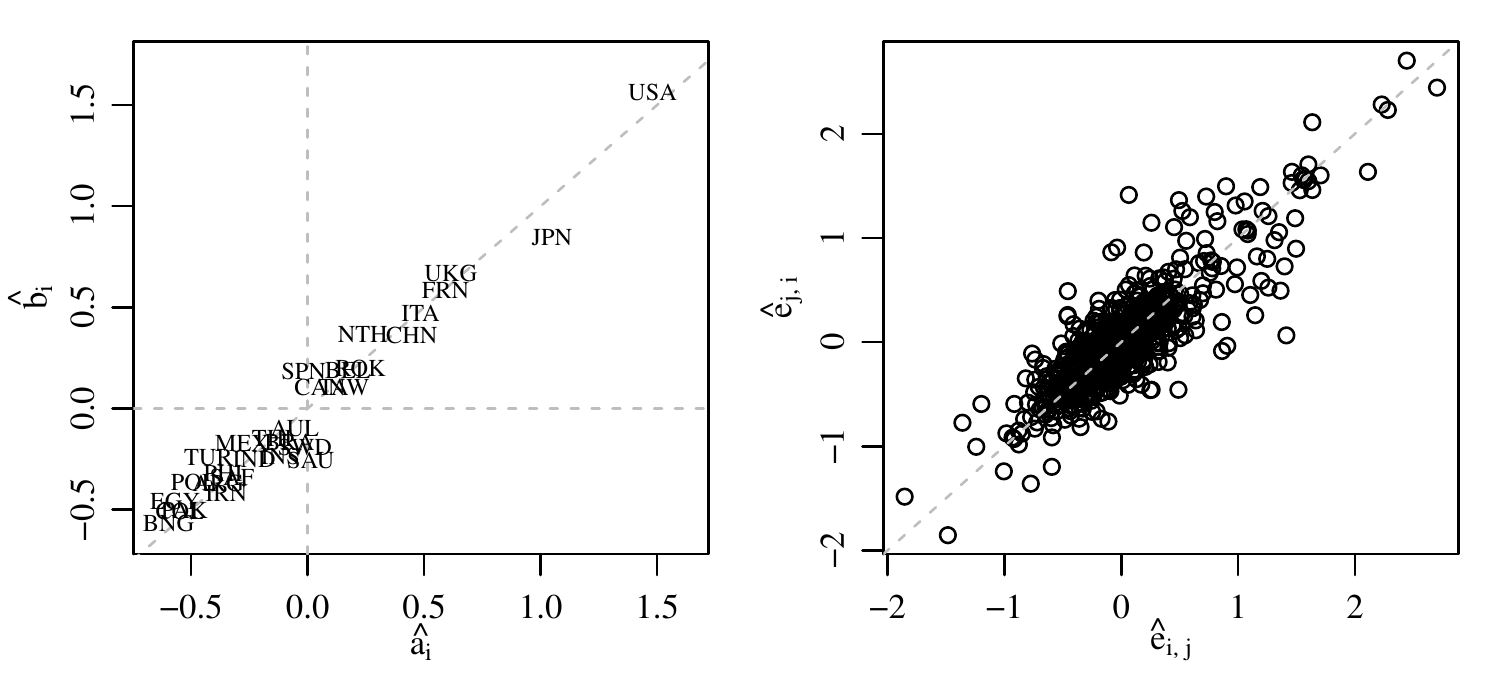} 

}

\end{knitrout}

\caption{Left panel: Scatterplot of country-level export 
effects versus import effects. Right panel: Scatterplot of 
dyadic residuals.}
\label{fig:trade_eda}
\end{figure}

We illustrate these phenomena empirically with 
a sociomatrix
of export data among $n=30$ countries. Here, $y_{i,j}$ is the
1990 export volume 
from country $i$ to country $j$, 
in log billions of dollars. 
For each country $i=1,\ldots,n$,  $\hat a_i$  the  
$i$th row mean minus the grand mean $\hat \mu$ of the sociomatrix, 
and $\hat b_i$ is
the $i$th column mean minus $\hat\mu$. 
The left panel of Figure \ref{fig:trade_eda} shows that 
these row and column effects are strongly correlated  -
countries 
with large export volumes  typically
have larger than average import volumes as well. 
A scatterplot of 
$\hat \epsilon_{i,j} = y_{i,j} - (\hat\mu + \hat a_i + \hat b_j )$
versus  $\hat \epsilon_{j,i}$ in the right panel of the plot 
indicates a strong dyadic correlation, even 
after controlling for country-specific heterogeneity in export and import volumes. 

The standard ANOVA model of a data matrix quantifies  
row variation, column variation and residual variation.  
However, the ANOVA model does not quantify the sender-receiver or dyadic  correlations
that are apparent from the figure, 
and that are present in 
most other dyadic datasets I have 
seen.
A model that does  quantify these correlations, and therefore 
provides a more complete description of the sociomatrix, 
was introduced in the psychometrics literature 
by \citet{warner_kenny_stoto_1979}. 
This more complete model, called the social relations model (SRM), 
is a  random effects model
given 
by \ref{eqn:rce} but with the additional assumptions that 
\begin{equation} 
\Var{  (\begin{smallmatrix} a_i \\ b_i \end{smallmatrix} ) } = \Sigma = \begin{pmatrix} \sigma^2_a & \sigma_{ab} \\  
    \sigma_{ab} & \sigma_b^2 \end{pmatrix}   \ \ \ 
\Var{ ( \begin{smallmatrix} \epsilon_{i,j} \\ \epsilon_{j,i}\end{smallmatrix}) } =   \sigma^2 \begin{pmatrix}  1 & \rho \\ \rho & 1 \end{pmatrix},  
\label{eqn:srm}
\end{equation} 
with effects otherwise being independent. 
Straightforward calculations show that
under this random effects model, 
the variance of the relational variable is 
$\Var{ y_{i,j} } = \sigma^2_a + 2 \sigma_{ab} + \sigma^2_b + \sigma^2$, 
and the covariances among the relations are 
\begin{align*} 
\Cov{ y_{i,j} ,y_{i,k} } & =  \sigma_a^2  & \text{(within-row covariance)}   \\ 
\Cov{ y_{i,j} ,y_{k,j} } & =  \sigma_b^2 & \text{(within-column covariance)}  \\
 \Cov{ y_{i,j} ,y_{j,k} } & =  \sigma_{ab} & \text{(row-column covariance)}     \\
 \Cov{ y_{i,j} ,y_{j,i} } & =  2 \sigma_{ab} + \rho \sigma^2 & \text{(row-column covariance plus reciprocity)}  
\end{align*}
with all other covariances between elements of $\bl Y$ being zero.
We refer to this covariance model as the \emph{social relations covariance
model}.  Unbiased moment-based   
estimators of $\mu$, $\Sigma$, $\sigma^2$ and $\rho$
are derived in \citet{warner_kenny_stoto_1979},  and
standard errors for these estimators are obtained in 
\citet{bond_lashey_1996}. Under the additional assumption 
that the random effects are jointly normally distributed, 
\citet{wong_1982} provides an EM algorithm for maximum 
likelihood estimation, \citet{gill_swartz_2001} develop a 
Bayesian method for parameter estimation, and
\citet{li_loken_2002} discuss connections to models in 
genetics and extensions to repeated-measures dyadic data.  

\subsection{Social relations regression models}

\begin{table}
\begin{tabular}{r|ccc|ccc|ccc}
&  \multicolumn{3}{c|}{IID} &  \multicolumn{3}{c}{SRRM} &  \multicolumn{3}{c}{AME} \\ \hline
regressor &   $\hat\beta$ & se($\hat\beta$)  & $t$-ratio &   $\hat\beta$ & se($\hat\beta$) & $t$-ratio  &  $\hat\beta$ & se($\hat\beta$)  & $t$-ratio  \\ \hline
exporter polity&   0.015 &   0.004 &   4.166 & 
                   0.015 &   0.016 &   0.934 & 
                   0.012 &   0.016 &   0.782  \\ 
importer polity&   0.022 &   0.004 &   6.070 &
                   0.022 &   0.016 &   1.419 &
                   0.018 &   0.015 &   1.190  \\ 
exporter GDP   &   0.411 &   0.021 &  19.623 &
                   0.407 &   0.095 &   4.302 &
                   0.346 &   0.103 &   3.373  \\ 
importer GDP   &   0.398 &   0.020 &  19.504 &
                   0.397 &   0.094 &   4.219 &
                   0.336 &   0.103 &   3.250  \\ 
distance       &  -0.057 &   0.004 & -13.360 &
                  -0.064 &   0.005 & -11.704 &
                  -0.041 &   0.004 & -10.970  \\ 
\end{tabular}
\caption{Parameter estimates and standard errors from the trade data using a normal linear regression model with i.i.d.\ errors, a SRRM, and an AME model.}
\label{tab:trade_srm}
\end{table}

Often we wish to quantify the association between
a particular dyadic variable and
some other dyadic or nodal variables. 
Useful for such situations is a type of linear mixed effects
model we refer to as the \emph{social relations regression model} (SRRM),
which combines a linear regression model with the covariance
structure of the SRM as follows:
\begin{equation}
 y_{i,j} = \bs \beta^\top \bl x_{i,j} +  a_i + b_j +  \epsilon_{i,j} ,  
\label{eqn:srrm}
\end{equation} 
where $\bl x_{i,j}$ is a $p$-dimensional vector of  regressors
and $\bs\beta$ is a vector of regression coefficients 
to be estimated. 
The vector $\bl x_{i,j}$ may contain variables that are specific to 
nodes or pairs of nodes. For example, we may have 
$\bl x_{i,j} = (  \bl x_{r,i}, \bl x_{c,j}, \bl x_{d,i,j})$ where
$\bl x_{r,i}$ is a vector of characteristics
of node $i$ as  a sender or row object,
$\bl x_{c,j}$ is a vector of characteristics of node $j$
as a receiver or column object, 
and $\bl x_{d,i,j}$ is a vector of characteristics of the ordered pair $(i,j)$.

We illustrate the use of the SRRM with a more detailed analysis of 
the international trade dataset described above. This dataset also includes
several other variables, such as  country-specific measures
of gross domestic product (GDP) and polity (a measure of
citizen access to government), as well as the geographic distance between
pairs of county capitals. Our objective in this example is 
to quantify the relationship between trade and polity after
controlling for the effects of GDP and geographic distance.
We first do so with
a naive ordinary linear regression model 
of the form
\[
  y_{i,j} = \beta_0 + \beta_{r,1} \text{polity}_{i} + 
             \beta_{r,2} \text{gdp}_{i} +  
             \beta_{c,1} \text{polity}_{j} + 
             \beta_{c,2} \text{gdp}_{j}  +  
             \beta_{d} \text{distance}_{i,j} +  \epsilon_{i,j}, 
\] 
where $\text{polity}_i$ is a measure of country $i$'s polity score on 
a scale from 1 to 10, 
$\text{gdp}_i$ is the log GDP of country $i$ in dollars, 
 $\text{distance}_{i,j}$ is the log  distance  in miles between capitals 
of countries $i$ and $j$, and 
the $\epsilon_{i,j}$'s are assumed to be i.i.d.\ mean-zero error terms. 
This model is  a ``gravity model''  of trade 
\citep{isard_1954,bergstrand_1985}, where trade flow is analogous to 
a gravitational force between countries, and GDP plays the role of 
mass.
Gravity models of this type are 
widely used to empirically evaluate different theories of international 
trade
\citep{baier_2009}.

Regression parameter estimates and standard errors
assuming an i.i.d.\ error model
are given in  the first column of Table \ref{tab:trade_srm}. 
Based upon the ratio of parameter estimates to standard errors, 
we would conclude that the hypothesis of no polity effects is inconsistent 
with an i.i.d.\ error model. 
However, while technically valid, this conclusion is not particularly 
interesting given that we 
expect row, column and dyadic 
dependence for network data such as these, 
and thus doubt the i.i.d.\ error model {\it a priori}. 
 More interesting is 
an evaluation of  whether or not the hypothesis of no polity effects  
is consistent with a social relations covariance model. The parameter 
estimates and standard errors for the SRRM in the second column of the table indicate that indeed it is: 
the parameter estimates of the polity effects are not substantially 
larger than their standard errors.

\begin{figure}
\begin{knitrout}\footnotesize
\definecolor{shadecolor}{rgb}{0.969, 0.969, 0.969}\color{fgcolor}

{\centering \includegraphics[width=6in]{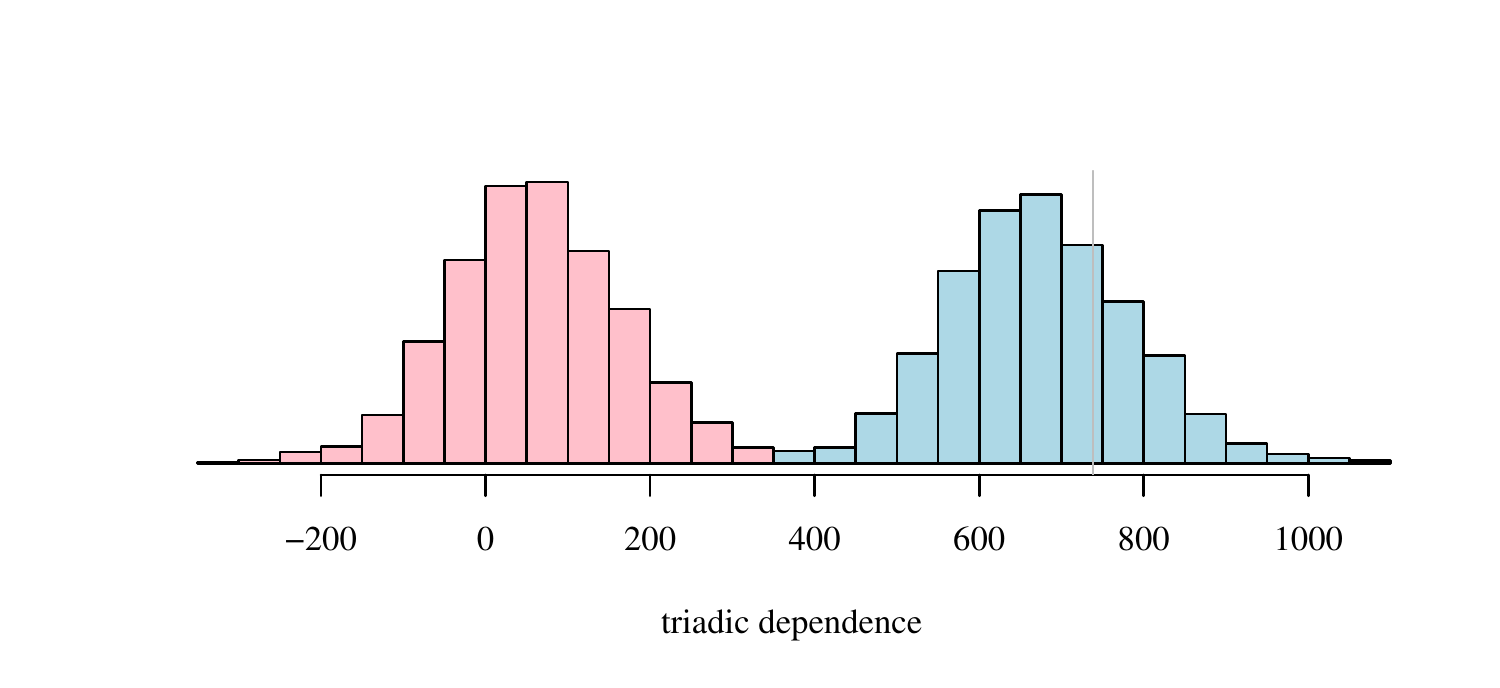}

}

\end{knitrout}
\caption{Posterior predictive distributions
of a triadic goodness of fit statistic.
The pink histogram corresponds to the SRRM fit, the
blue to the AME fit. The observed value of the statistic is given by
the vertical gray line. }
\label{fig:trade_gof}
\end{figure}

\section{Multiplicative Effects  Models}  
While more reasonable than an ordinary regression model, 
SRRMs applied to many datasets often exhibit substantial lack 
of fit. In particular, it is often observed that real networks 
exhibit patterns of dependence among triples of nodes 
such as transitivity, balance and clustering 
\citep{wasserman_faust_1994}.
For example, in the context of fitting a regression model, 
the notion of balance  would correspond to there generally being a 
higher-than expected 
relationship (i.e.\ a positive residual) between nodes $j$ and $k$ if that between 
$i$ and $j$ and $i$ and $k$ were both also higher than expected. 
Such patterns can be quantified with summary statistics such 
as $\sum_{i,j,k} \hat\epsilon_{i,j} \hat\epsilon_{j,k} \hat\epsilon_{k,i}$, 
where $\hat \epsilon_{i,j}$ is a residual from a least-squares fit. 
Figure \ref{fig:trade_gof} displays the posterior predictive distribution 
of this statistic from a Bayesian fit of the SRRM to the trade data. 
The predictive distribution of this statistic under the SRRM 
does not overlap with the observed value, indicating that 
the SRRM is inconsistent with this feature of the data. 

The SRRM, or any other Gaussian random effects model, is unable 
to  describe a third-order dependence pattern such as this
because all third-order moments of mean-zero Gaussian random variables 
are zero. To model such patterns we must move beyond linear Gaussian 
random effects models where random effects 
and error terms combine additively. One solution
is to consider additional random effects
 that combine 
nonadditively. For example, let $\gamma_{i,j} = \bl u_i^\top \bl v_j$, 
where
$\bl u_i$ and $\bl v_i$ are $r$-dimensional 
 mean-zero latent Gaussian  vectors, i.i.d.\ across nodes with 
$\Cov{ \bl u_i , \bl v_i} = \Psi_{uv}$. 
Then
\begin{align*}
\Exp{ \gamma_{i,j} \gamma_{j,k} \gamma_{k,i} } &=  
\Exp{  \bl u_i^\top\bl  v_j \bl u_j^\top\bl  v_k \bl u_k^\top \bl  v_i}  \\
& =  \Exp{ \bl u_i^\top \bl v_i }^3  
 = \text{tr}(\Psi_{uv})^3. 
\end{align*} 
Therefore, a social relations regression model that includes 
second- and third-order residual dependencies 
is given by 
\begin{align}
y_{i,j} & = \bs\beta^\top \bl x_{i,j}  + 
  \bl u_i^\top \bl v_j + a_i + b_j + \epsilon_{i,j}   \label{eqn:game}  \\
(\bl u_1,\bl v_1) ,\ldots, (\bl u_n, \bl v_n) & \sim \text{i.i.d.} \
  N_{2r}(\bl 0, \Psi )   \nonumber \\
(a_1,b_1) ,\ldots, (a_n,b_n) & \sim \text{i.i.d.}  \ N_2(\bl 0, \Sigma)  \nonumber \\
\{ (\epsilon_{i,j}, \epsilon_{j,i} ) : i<j \} & \sim \text{i.i.d.}  \
  N_2( \bl 0 , \sigma^2  (\begin{smallmatrix}
       1 & \rho \\ \rho & 1 \end{smallmatrix} ) ) . \nonumber
\end{align}
We call such  a model an \emph{additive and multiplicative effects} 
model (AME). Specifically, we refer to the model given 
by (\ref{eqn:game}) as a Gaussian AME, since the observed data 
are conditionally Gaussian, given $\bs\beta$ and the multiplicative 
effects. 
A rudimentary multiplicative effects model appeared 
in \citet{hoff_raftery_handcock_2002}, along with 
some other nonadditive random effects models.
A symmetric multiplicative effects model 
was combined with the social relations covariance model in 
\citet{hoff_2005}, and versions of  (\ref{eqn:game})
were studied and developed in 
\citet{hoff_2008,hoff_2009} and \citet{hoff_fosdick_volfovsky_stovel_2013}. 

The matrix form of this model can be expressed as 
\[
\bl Y=\bl M + \bl a\bl 1^\top  + \bl 1 \bl b^\top + \bl U \bl V^\top + \bl E,
\]
where $m_{i,j} = \bs\beta^\top \bl x_{i,j}$,   
$\bl a= (a_1,\ldots, a_n)$, $\bl b= (b_1,\ldots, b_n)$ 
and $\bl U$ and $\bl V$ are $n\times r$ matrices 
with $i$th rows equal to $\bl u_i$ and $\bl v_i$ 
respectively, with $r$ being the length of each of these latent 
vectors.  
This represents the deviations of $\bl Y$ 
from the linear regression model $\bl M$
as the sum of 
a rank-1 matrix of 
row effects, a rank-1 matrix of 
column effects, 
a rank-$r$ matrix $\bl  U\bl V^\top$ and a noise matrix $\bl E$. 
Absent a covariance model for the node-specific effects 
or dyadic residuals, this representation is essentially 
a special case of an  
\emph{additive main effects, multiplicative interaction} (AMMI) model
\citep{gollob_1968, bradu_gabriel_1974}, a class of
matrix models developed in the psychometric and agronomy literature
for data arising from two-way layouts with no replication. 
Since sociomatrices have additional structure - 
the row factors are the same as the column factors - 
our random effects version of the AMMI model 
includes the SRM covariance model
for the $a_i$'s, $b_i$'s and $\epsilon_{i,j}$'s, in addition to a random effects model
for $\bl u_i$ and $\bl v_i$ to represent possible 
third-order dependencies in the sociomatrix. 
We refer to this model as an \emph{additive and multiplicative effects} model, or
AME model for dyadic network data.

To illustrate the how the inclusion of multiplicative effects improves model 
fit, we obtain the posterior predictive distribution of the 
triadic goodness-of-fit statistic
$\sum_{i,j,k} \hat\epsilon_{i,j} \hat\epsilon_{j,k} \hat\epsilon_{k,i}$
under an AME model with two-dimensional multiplicative
effects and 
the same regressors as the SRRM  (polity, GDP and geographic distance). 
A histogram of this posterior predictive distribution is given 
in  Figure \ref{fig:trade_gof}, along with that of the 
SRRM fit. The posterior predictive distribution obtained 
under the AME fit is roughly centered around the observed 
value of the statistic indicating that, unlike the SRRM, 
the AME model is able to describe this third-order residual dependency in the 
trade data. Finally, parameter estimates and standard errors for the regression 
coefficients in this AME model are given in the third column of 
Table \ref{tab:trade_srm}.  Parameter estimates are slightly smaller than 
those of the SRRM, but the main conclusions remain the same.

\begin{figure}
\begin{knitrout}\footnotesize
\definecolor{shadecolor}{rgb}{0.969, 0.969, 0.969}\color{fgcolor}

{\centering \includegraphics[width=6in]{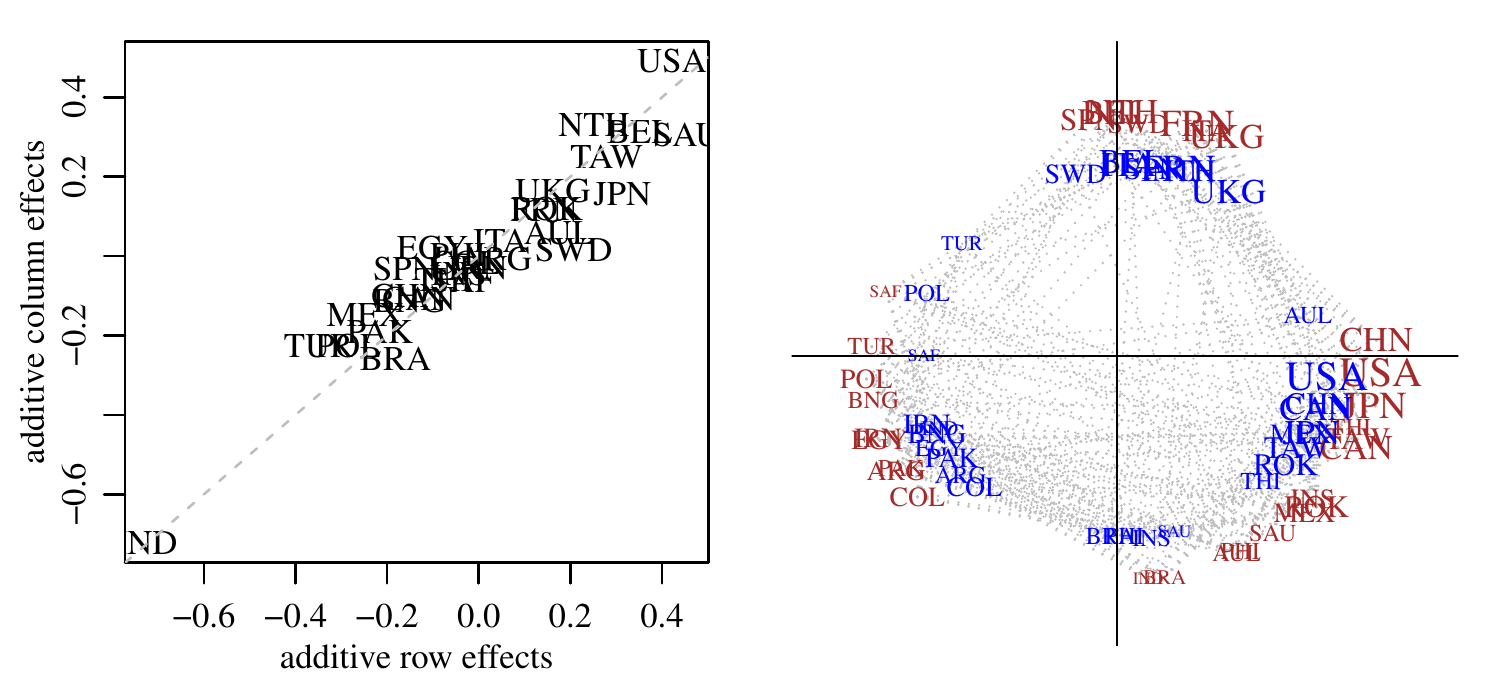} 

}

\end{knitrout}
\caption{Estimates of node-specific effects. The left panel 
gives  additive row effects versus additive column effects. 
The plot on the right gives estimates of $\bl u_i$ in red
and $\bl v_i$ in blue for each country $i=1,\ldots, n$.  
The country names indicate the direction of these vectors, 
and the size of the plotting text indicates their magnitude. 
A dashed line is drawn between an export-import pair if 
their trade flow is larger than expected based on 
the other terms in the model.}
\label{fig:lvplot}
\end{figure}

From a random effects perspective, the multiplicative 
effect $\bl u_i^\top \bl v_j$ can 
be viewed as a means to quantify third 
order dependence. However, these effects can also be 
interpreted as representing omitted regression variables 
or uncovering group structure among the nodes. 
This interpretation
is based on the observation 
that the 
strength or presence of ties
between 
  nodes 
is often related to similarities
of node-level attributes.
For example, suppose  for each node $i$  that $x_i$
is the indicator that $i$
is a member of a particular  group or  has a particular trait.
Then $x_{i}x_{j}$ is the indicator that $i$ and $j$ are
co-members of this group, and this fact may have
some effect on their relationship $y_{i,j}$.
A positive association between $x_ix_j$ and $y_{i,j}$ is
referred to as homophily, and a negative association
as anti-homophily.
Quantifying homophily on an observed
attribute can be done with a SRRM
by creating a dyadic regressor  $x_{d,i,j}$
from a nodal regressor $x_i$ through multiplication
($x_{d,i,j} = x_i x_j$) or some other operation.
However, the possibility that not all relevant nodal 
attributes 
are included in a network dataset motivates inclusion 
of the multiplicative term $\bl u_i^\top \bl v_j$, 
where $\bl u_i$  and $\bl v_i$ represent unobserved latent factors of 
node $i$ as a sender and receiver of relations, respectively. 

These latent factors may be estimated  and examined to 
highlight additional structure in the data beyond 
that explained by the SRRM. For example, estimates of 
the $\bl u_i$'s and $\bl v_i$'s of the rank-2 AME fit to the trade 
data are displayed in Figure \ref{fig:lvplot}. 
Recall that this model includes polity, GDP and 
geographic distance as regressors, in addition to 
the additive effects and multiplicative latent factors. 
The interpretation of the multiplicative factors is that if $\bl u_i$ and 
$\bl v_j$ are large and in the same direction, then 
nodes $i$ and $j$ tend to have observed trade flows 
larger than $\bs\beta^\top \bl x_{i,j} + a_i +b_j$, that is, 
larger than what is predicted by the additive effects alone. 
As can be seen from the figure, 
the estimates of the latent factors from  these data 
highlight some geographically related clustering of 
nodes, in particular,  a cluster of Pacific rim countries and a cluster 
of mostly European countries. These are patterns that, while 
related to geographic distance, are not well-represented by 
a single linear relationship between log-trade and 
log-distance in the regression model.

\section{Transformation models for non-Gaussian networks}
On their original scale, many dyadic variables are not well-represented 
by a model with Gaussian errors. 
In some cases, such as with 
the trade data, a dyadic  variable can be transformed
so that the
Gaussian
AME model is reasonable.  In other cases,
such as with binary, ordinal, discrete or sparse variables,
 no such transformation is available.
Examples of such data include measures of friendship that
are binary (not friends/friends) or ordinal (dislike/neutral/like),
discrete counts of conflictual events between countries, or the amount of time
two people spend on the phone with each other. 
In this section we describe extensions of the Gaussian AME model
to accommodate ordinal dyadic data, where in what follows,
ordinal means any outcome for which the possible values
can be put in some meaningful order. This includes discrete outcomes
(such as binary indicators or counts), ordered qualitative
outcomes (such as low/medium/high),
and even continuous outcomes.
The extensions are based on latent variable
representations of probit and ordinal probit regression models.

\subsection{Binary and ordinal network data} 
Let $\bl S$ be the observed sociomatrix for a dyadic variable $s_{i,j}$.
The simplest type of ordinal dyadic variable is a binary variable indicating the presence  of some
type of relationship between $i$ and $j$,
so that $s_{i,j} = $ 0 or 1 depending on whether a social 
link  is absent or present, respectively.
One approach to quantifying the association between such a  binary  
variable and other variables is with probit 
regression, which models the probability of  
a link between $i$ and $j$ as $\Phi( \bs\beta^\top \bl x_{i,j})$, 
where $\Phi$ is the standard normal CDF. As is well known, the 
probit regression model has a latent variable representation 
in which $s_{i,j}$ is the binary indicator that 
some latent normal random variable $y_{i,j}\sim N(\bs\beta^\top \bl x_{i,j},1)$
is greater than zero
\citep{albert_chib_1993}. 
An ordinary probit regression model corresponds to the 
$y_{i,j}$'s being independent, which is generally an inappropriate 
assumption for  
network data. However, 
a model for binary data that does capture 
the types of network dependencies discussed 
in the previous section, such as row and column covariance, 
dyadic correlation, and triadic dependence, can be represented 
via an AME model for the latent $y_{i,j}$'s: 
\begin{align} 
\label{eqn:tame}
y_{i,j} & = \bs\beta^\top \bl x_{i,j} +  \bl u_i^\top \bl v_j + a_i + b_j + 
   \epsilon_{i,j} \\ 
s_{i,j} &= g( y_{i,j} ),  \nonumber 
\end{align}
where the $a_i$'s $b_i$'s and $\epsilon_{i,j}$'s follow the 
SRM covariance model and 
$g(y)$ is the binary indicator that 
$y>0$.  Absent the multiplicative term 
$\bl u_i^\top \bl v_j$, this is basically  a 
generalized linear mixed effects model. Including the multiplicative term 
but absent the SRM covariance structure, this model is a type 
of generalized bilinear regression \citep{gabriel_1998}.  
Including both the multiplicative term and the SRM covariance structure 
yields a regression model for binary social network data that 
can accommodate second- and third-order dependence patterns. 

This probit AME model
for binary data
extends in a natural way to accommodate
ordinal data with more than two levels.
As with binary data, we model the
observed sociomatrix $\bl S$ as being
a function of a latent sociomatrix $\bl Y$
that follows a Gaussian AME distribution. Specifically, 
the model is the same as in Equation \ref{eqn:tame} 
but with
$g$ being a non-decreasing function.
Such a model may be viewed as a type of Gaussian transformation model
\citep{bickel_ritov_1997}. 

One approach to estimation for these models is as follows: 
For both the probit and ordinal probit models, observation of 
$\bl S$ tells us that $\bl Y$ lies in a certain set, 
say $\bl Y \in C(\bl S)$. For the binary probit model, this 
set is simply  given by $C(\bl S ) = \{ \bl Y\in \mathbb R^{n\times n} : 
   \text{sign}(y_{i,j}) =  \text{sign}( 2 s_{i,j}-1 )\}$, that is, 
$s_{i,j}=1$ implies $y_{i,j}>0$ and $s_{i,j}=0$ implies 
$y_{i,j}<0$. For the ordinal probit model, 
since $g$ is non-decreasing we have 
$C(\bl S) =  \{ \bl Y\in \mathbb R^{n\times n}:  
   \max_{i'j'} \{ y_{i',j'} : s_{i',j'}< s_{i,j} \}   < 
   y_{i,j} < \min_{i'j'} \{ y_{i',j'} : s_{i,j}< s_{i'j'}\} \}$. 
A likelihood based on the  knowledge that $\bl Y \in C(\bl S)$ is 
given by $L(\bs\theta) = \Pr( \bl Y\in C(\bl S) | \bs\theta )$ 
where $\bs\theta$ are the parameters in the Gaussian AME model 
for $\bl Y$.  While a closed form expression for this likelihood 
is unavailable, a Bayesian approach to estimation and inference is 
feasible via Gibbs sampling by iteratively simulating  
$\bs\theta$ from its full conditional distribution given $\bl Y$, then 
simulating 
$\bl Y$ from its conditional distribution given $\bs\theta$ but  
constrained to 
lie in $C(\bl S)$. 
More details are presented in Section 6.

\subsection{Censored and ranked nomination data}
Data on human social networks are often
obtained by asking
participants in a study to name and rank
a fixed number of people with whom they are friends. 
Such a survey method is called a \emph{fixed ranked nomination} (FRN)
scheme, and is used in studies of institutions such as
schools or businesses. For example,
the National Longitudinal Study of Adolescent Health (AddHealth,
\citet{harris_2009}) asked middle and high-school students to nominate and
rank up to five members of the same sex as friends, and five members
of the opposite sex as friends.

Data obtained from FRN schemes are similar to ordinal data,
in that the ranks of a person's friends may be viewed as
an ordinal response. However, FRN data are also censored
in a complicated way.
Consider a study where people were asked to name and rank up to
and including their top five friends. If person $i$
nominates five people but doesn't nominate person $j$, then $s_{i,j}$ is
censored: The data cannot tell us whether $j$ is $i$'s sixth best
friend, or whether $j$ is not liked by $i$ at all.
On the other hand, if person $i$ nominates four people as friends
but could have nominated five, then
person $i$'s data are not censored -  the absence of a nomination
by $i$ of $j$ indicates that $i$ does not consider $j$ a friend.

A likelihood-based approach to modeling
FRN data using an AME model was developed in
\citet{hoff_fosdick_volfovsky_stovel_2013}.
Similar to the approach for ordinal dyadic data described above,
this methodology treats the observed ranked outcomes $\bl S$
as a function of an underlying continuous sociomatrix $\bl Y$ of affinities
that
is generated from an
AME model.
Letting $m$ be the maximum number of nominations allowed, and
coding $s_{i,j} \in \{ m,m-1,\ldots, 1,0\} 
$ so that $s_{i,j}=m$ indicates that $j$ is $i$'s most liked friend,
the FRN likelihood is derived from the following constraints that the
observed ranks $\bl S$ tell us about the underlying dyadic variables $\bl Y$:
\begin{eqnarray}
s_{i,j}> 0 &\Rightarrow & y_{i,j}>0  \label{eqn:porc}\\
s_{i,j}  > s_{i,k} & \Rightarrow &  y_{i,j} > y_{i,k} 
  \label{eqn:rnkc}\\ 
s_{i,j} = 0  \ \mbox{and} \  
 d_i < m &\Rightarrow & 
    y_{i,j}\leq 0  \label{eqn:degc}. 
\end{eqnarray}
Constraint (\ref{eqn:porc})
indicates that
if $i$ ranks $j$, then $i$ has a positive relation with $j$ ($y_{i,j}>0$),
and constraint (\ref{eqn:rnkc})  indicates that
a higher rank corresponds to a more positive relation.
Letting $d_i\in \{ 0,\ldots, m\}$ be the number of people
that $i$ ranks, constraint (\ref{eqn:degc})
indicates that if
$i$ could have made additional friendship nominations
but chose not to nominate $j$,
they then do not consider $j$ a friend.
However, if
$s_{i,j}=0$ but $d_i = m$ then person $i$'s unranked relationships
are censored, and so $y_{i,j}$ could be positive
even though
$s_{i,j}=0$.  In this case, all that is known about
$y_{i,j}$ is that it is less than $y_{i,k}$ for any person
$k$ ranked by $i$. In summary, observation of $\bl S$ tells us 
that $\bl Y \in C(\bl S)$ where $C(\bl S)$ is defined by 
conditions \ref{eqn:porc} - \ref{eqn:degc}. 
As with the probit and ordinal AME models, Bayesian inference for 
this transformation model can proceed by 
iteratively simulating values of the model parameters and the 
unknown values of $\bl Y$ from their full conditional distributions. 

\section{Comparisons to other models} 
Two popular categories of statistical network models are 
exponentially parameterized random graph models (ERGMs) and 
latent variables models.  
Roughly speaking, ERGMs focus on characterizing global, macro-level patterns in a
network, while latent variable models describe  local, micro-level patterns of 
relationships among specific nodes. 
The AME class of model can characterize both global and 
local patterns, the former via the global parameters 
$\{ \bs\beta, \Sigma, \Psi, \sigma^2,\rho  \}$ and 
the latter via the node-specific factors $\{ a_i,b_i , \bl u_i, \bl v_i: i=1,\ldots, n\}$. 

\subsection{Comparisons to ERGMs}
An ERGM is a probability model for a binary sociomatrix that 
includes densities of the form 
$p( \bl  Y) = c(\bs\theta) \exp(\bs\theta\cdot \bl t(\bl Y )) $, 
where $\bl t(\bl Y)$ is a vector of sufficient statistics and 
$\bs\theta$ is a parameter to be estimated. 
Typical applications  
use a small number of sufficient statistics, 
often much smaller than the number of nodes, and in this sense 
the models describe ``global'' patterns in the data. 
An exception to this is the not-infrequent inclusion of 
out- and in-degree statistics that can characterize the 
differential sociability and popularity of the nodes. 
For example, one of the first ERGMs to be widely used and studied was 
the ``$p_1$'' model \citep{holland_leinhardt_1981} with density
\[
 p(\bl Y) \propto \exp\left ( \mu \sum_{i,j} y_{i,j} + 
    \sum_i (a_i \sum_j y_{i,j}  + b_i \sum_j y_{j,i} ) +
    \rho  \sum_{i,j} y_{i,j} y_{j,i} \right ), 
\]
which includes as sufficient statistics the total number of 
ties $\sum_{i,j} y_{i,j}$, the number of reciprocated ties 
 $\sum_{i,j} y_{i,j} y_{j,i}$ and the in- and out-degrees
 $\{\sum_j y_{i,j}, \sum_j y_{j,i}, i=1,\ldots,n\}$. 
The parameters in this model represent roughly the same data 
features as they do in the SRM: 
an overall mean of the relations ($\mu$), 
heterogeneity in row and column means (the $a_i$'s and $b_i$'s)
and dyadic correlation ($\rho$). 
Similarities are also found between the SRRM and the 
``$p_2$'' model 
developed by
\citet{vanduijn_snijders_zijlstra_2004}. The $p_2$ model  
extends the $p_1$ model by including regressors
(as does the SRRM), and additionally 
treats the node-level parameters $a_i$ and $b_i$ as 
potentially correlated random effects (as do the SRM and SRRM). 

\citet{holland_leinhardt_1981} concede that the $p_1$ model 
is of limited utility due to its inability to describe more 
complex forms of dependency such as transitivity or clustering. 
While inclusion of appropriate regressors, either in 
a $p_2$  model or SRRM, can represent some degree of  higher-order 
dependency, often such models still exhibit lack-of-fit 
and more complex models are desired. 
As described in Section 3, the AME approach
is to include a multiplicative latent variable term $\bl u_i^\top \bl v_j$, 
that when thought of as a random effect, 
induces non-zero third order moments in the error structure. 
In contrast, 
the ERGM approach to describing higher-order dependencies is to include 
additional sufficient statistics, such as the number of triangles observed 
in the graph, or the number of cycles of various lengths 
\citep{snijders_2006}. Unfortunately, simultaneous inclusion of such 
statistics and those that naturally represent degree heterogeneity can 
lead to model degeneracy  
\citep{handcock_2003,hunter_handcock_2006}. 

\subsection{Comparison to other latent variable models}
While globally inducing third-order dependence among network 
outcomes, the multiplicative term $\bl u_i^\top \bl v_j$ in the AME model
can also
 be interpreted locally at the micro-level, in that 
$\bl u_i$  and $\bl v_i$ describe latent features of node $i$ as a sender 
and receiver of ties. Estimates of the features (such as those displayed in 
 Figure \ref{fig:lvplot}) can be used to identify interesting nodes, 
 assist with visualization of network patterns, or be used as an 
input to other data analysis methods, such as 
clustering 
\citep{rohe_chatterjee_yu_2011}. 
Other popular non-additive  latent variable models for network data 
include the stochastic blockmodel  \citep{nowicki_snijders_2001} and 
the latent distance model  \citep{hoff_raftery_handcock_2002}. 
The blockmodel assumes each node belongs to 
an unobserved latent class or ``block'', and that the relations 
between two nodes are determined (statistically) by their 
block memberships. This model is based on the assumption of 
\emph{stochastic equivalence}, that is, the assumption that 
the nodes 
can be divided into groups such that members of the same group have 
the same distribution of 
relationships to other nodes. 
In contrast, the distance model assumes each node has some unobserved 
location in a latent ``social space,''  and that the strength of 
a relation between two nodes is decreasing in the distance between 
them in this space.  This model provides a compact representation of 
certain patterns seen in social networks such as 
transitivity and community, that is, the existence subgroups of nodes with strong within-group relations.

\begin{figure}
\begin{knitrout}\footnotesize
\definecolor{shadecolor}{rgb}{0.969, 0.969, 0.969}\color{fgcolor}

{\centering \includegraphics[width=6in]{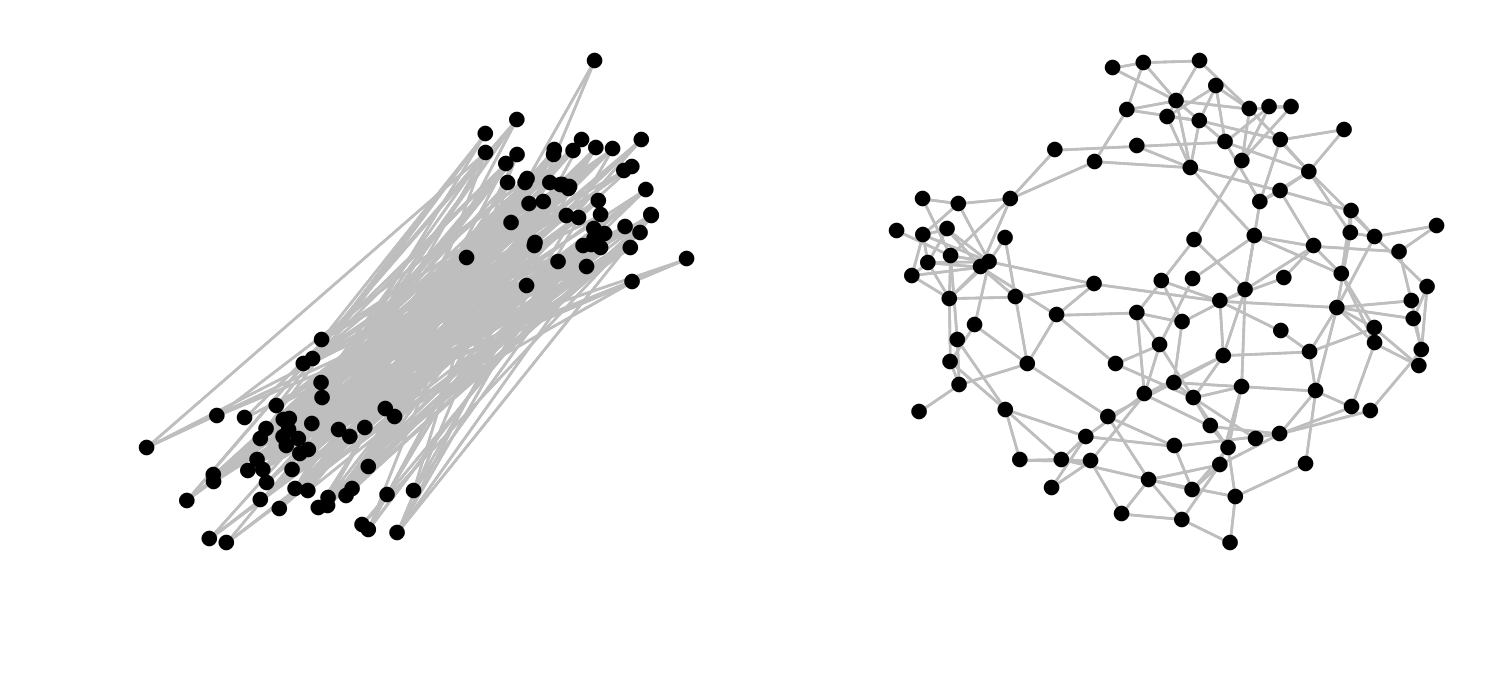}

}

\end{knitrout}
\caption{Two hypothetical networks. The network on the left 
can be represented by two 
groups of stochastically equivalent nodes. The network on the 
right can be represented by an embedding of the nodes in 
two-dimensional Euclidean space. }
\label{fig:exnet}
\end{figure}

Figure \ref{fig:exnet}  displays two hypothetical symmetric networks, 
each one of which can be well-represented by one of these 
two latent variable models. 
The network on the left can be well-represented by 
a two-group stochastic blockmodel in which the within-group 
density of ties is lower than the between-group density. 
Such a network is not representable by a latent distance model
because in such a model, stochastic equivalence of two nodes 
is confounded with the expected strength of their relationship:
In a latent distance model, two nodes are stochastically 
equivalent if they are in the same location in the social 
space. However, if they are in the same location, then the
distance between them is zero and
so their expected relationship is strong.
As such, networks 
where stochastically equivalent nodes have weak ties 
 will not be well-represented by a latent distance model. 
Conversely, the network displayed on the right side of 
Figure \ref{fig:exnet} is very well represented by 
a two-dimensional latent distance model 
in which the probability of a tie between two nodes 
is decreasing in the distance between them. 
However, representation of this network by a blockmodel 
would require a large number of blocks (e.g.\ one block in 
each subregion of the space), 
none of which would be particularly cohesive or distinguishable from 
the others. 

In contrast to these two extreme networks, 
real networks exhibit combinations of stochastic equivalence and 
transitivity in varying amounts. Inference based on either  a 
blockmodel or a distance model would then provide only an incomplete 
description of the heterogeneity across nodes in terms of how they 
form ties to others. 
Fortunately, as shown in \citet{hoff_2008}, latent variable models 
based on multiplicative effects (such as AME models) 
can represent both of these 
types of network patterns, and therefore provide a generalization 
of both the stochastic blockmodel and the latent distance model. 
To explain this generalization, we consider the simple case 
of an undirected dyadic variable so that the sociomatrix is symmetric. 
Each of the three types of latent variable models 
may be written abstractly as  
$y_{i,j}  \sim  m_{i,j} + \alpha(\bl u_i,\bl u_j)$
where $\alpha$ is some function of the node-specific latent 
variables $\bl u_1,\ldots, \bl u_n$, $m_{i,j}$ consists of any other 
terms in the model (such as a regression term or additive effects), 
and ``$y \sim x $'' means that the distribution of $y$ is 
stochastically increasing in $x$. 
The three latent variable models correspond to the following 
three specifications of the function $\alpha$:
\begin{description} 
\item{Stochastic blockmodel:} $\alpha(\bl u_i,\bl u_j)= \bl u_i^\top \Theta \bl u_j $, 
where $\bl u_i\in \mathbb R^r$ is a standard basis vector 
indicating block membership, and 
$\Theta$ is $r\times r$ symmetric.  
\item{Latent distance model:} $\alpha(\bl u_i,\bl u_j) = - |\bl  u_i - \bl u_j|$, where
 $\bl u_i  \in \mathbb R^r$. 
\item{Multiplicative effects model:} $\alpha(\bl u_i,\bl u_j) =\bl  u_i^\top \Lambda\bl  u_j$, 
where 
$\bl u_i  \in \mathbb R^r$ and $\Lambda$ is an $r\times r$ diagonal matrix. 
\end{description}
\citet{hoff_2008} referred to the symmetric multiplicative effects 
model as an ``eigenmodel'', as the 
matrix $\bl U\Lambda \bl U^\top$  resembles an 
eigendecomposition  of a rank-$r$ matrix. 
Note that as the $\bl u_i$'s range over $r$-dimensional 
Euclidean space, and $\Lambda$ ranges over all $r\times r$ 
diagonal matrices, the matrix 
$\bl U\Lambda\bl U^\top$ ranges over the space of all 
symmetric rank-$r$ matrices. 
Similarly, for the asymmetric AME models discussed elsewhere in this article,
as the $\bl u_i$'s and $\bl v_i$'s range over $r$-dimensional 
space, the multiplicative term $\bl U\bl V^\top$ ranges over the 
space of all  $n\times n$ rank-$r$ matrices. 

To compare these  models we compare the 
sets of  matrices that are representable by 
their latent variables. 
Let $\mathcal S_n$ be the 
set of $n\times n$ symmetric matrices, and let 
\begin{align*}
\mathcal B_r =& \{ \bl S \in \mathcal S_n: s_{i,j} = \bl u_i^\top \Theta\bl u_j,  
    \text{ $\bl u_i$ a standard basis vector }, 
        \text{ $\Theta\in \mathbb R^{r\times r}$ symmetric}\}; \\
\mathcal D_r =& \{ \bl S\in \mathcal S_n: s_{i,j} = -|\bl u_i-\bl u_j|,  \ 
       \bl u_i\in \mathbb R^r
  \};  \\
\mathcal E_r =& \{ \bl S\in \mathcal S_n: s_{i,j} = \bl u_i^T\Lambda \bl u_j, \ 
     \bl u_i \in \mathbb R^r, \ 
     \mbox{ $\Lambda$ a $r \times r $ diagonal matrix}\} .
\end{align*}
In other words, $\mathcal B_r$ is the set  of matrices 
expressible as a 
 $r$-dimensional  blockmodel, 
and $\mathcal D_r$ and $\mathcal E_r$ are defined similarly. 
\citet{hoff_2008} showed the following:
\begin{enumerate}
\item $\mathcal E_r$ generalizes $\mathcal B_r$;
\item $\mathcal E_{r+1}$ weakly generalizes $\mathcal D_r$;
\item $\mathcal D_{r}$ does not  weakly generalize $\mathcal E_1$. 
\end{enumerate}
Result 1 means that $\mathcal B_r$ is a proper subset of $\mathcal E_r$ 
unless $r\geq n$. This is  because the matrix $\bl S$  
corresponding to  an $r$-group
blockmodel is of rank $r$ or less, and $\mathcal E_r$ includes 
all such matrices. Result 2 means that for any  $\bl S\in  \mathcal D_r$, 
there exists an $\tilde {\bl S}\in \mathcal E_{r+1}$ 
whose elements are a monotonic transformation 
of those of $\bl S$, that is, have a numerical order that matches that of 
the elements of $\bl S$. 
Finally, result 3 says that there exist
rank-1 matrices $\bl S$, expressible via one-dimensional 
multiplicative effects, that cannot be order-matched by 
a distance model of \emph{any} dimension. 
Taken together, these results imply that 
multiplicative effects models 
can represent both the types of network patterns representable 
by  stochastic blockmodels and those representable by 
 latent distance models, and so is a more general 
and flexible class of models than either of these two 
other latent variable models. 
See \citet{hoff_2008} for more details and numerical examples. 

\section{Inference via posterior approximation}
While maximum likelihood estimation 
for  a Gaussian AME model is feasible, it is 
quite challenging for  binary, ordinal and other 
AME transformation models because the likelihoods involve
intractable integrals arising from the 
combination of the transformation and 
dependencies 
induced by the SRM. 
However, reasonably standard Gibbs sampling algorithms can be constructed 
to provide Bayesian inference 
for a wide variety of AME network models. 
We first construct a Gibbs sampler for Gaussian SRRMs,  
then extend the sampler to accommodate  Gaussian AME models, and 
finally extend the algorithm to fit AME transformation models. 
These algorithms are implemented in the 
{\sf R} package {\tt amen} \citep{amen}. 
\citet{hoff_2015} provides an {\sf R} vignette with several 
data analysis examples using these methods.

\subsection{Gibbs sampling for the SRRM} 
The unknown quantities in the Gaussian SRRM include the 
parameters $\bs\beta$, $\Sigma$, $\sigma^2$, and $\rho$, and the 
random effects $\bl a$ and $\bl b$.  
Posterior approximation for these quantities is 
facilitated by using 
 a $ N_p( \bs\beta_0 , \bl Q_0^{-1} )$ prior distribution for $\bs\beta$, 
a $\text{gamma}( \nu_0/2,\nu_0\sigma^2_0/2)$ prior 
distribution for $1/\sigma^2$ 
and a 
$\text{Wishart}( \Sigma^{-1}_0/\eta_0 ,\eta_0)$ prior 
distribution for $\Sigma^{-1}$. 
A Gibbs sampler proceeds by iteratively simulating the values of 
the unknown quantities from their conditional distributions, thereby 
generating a Markov chain having a stationary distribution equal 
to the  target posterior distribution. 
Values simulated from this Markov chain can be
used to approximate a variety of  posterior quantities of interest.
Given starting values of the unknown quantities, 
the algorithm proceeds
by iterating the following steps:
\begin{enumerate}
\item Simulate $\{ \bs\beta, \bl a, \bl b \}$   given
   $\bl Y$, $\Sigma$, $\sigma^2$, $\rho$;
\item Simulate $\sigma^2$  given
$\bl Y, \bs\beta, \bl a, \bl b, \rho$;
\item Simulate $\rho$  given
   $\bl Y, \bs\beta, \bl a, \bl b, \sigma^2$; 
\item Simulate $\Sigma$  given $\bl a,\bl b$; 
\item Simulate missing values of $\bl Y$ given 
    $\bs\beta, \bl a, \bl b, \sigma^2,\rho$  and  
     observed values of $\bl Y$. 
\end{enumerate}
We include the last step because, 
while sociomatrices typically
have undefined diagonals, 
 the calculations 
below make use of matrix operations that are only defined on 
matrices with no missing values.
By treating the diagonal values as 
missing at random, the fact that they are undefined will 
not affect the posterior distribution. Additionally, this 
step permits imputation of other dyadic outcomes that are 
missing at random.

Steps 2 through 5 are relatively standard. We discuss implementation 
of these steps before deriving the full conditional distribution of 
$\{ \bs\beta, \bl a, \bl b\}$.   
To implement steps 2 and 3, 
consider the stochastic representation of 
the SRRM as  
\begin{equation}
 \bl Y =  \bl M(\bl X, \bs\beta) + \bl a\bl 1^\top + \bl 1 \bl b^\top + \bl E 
\label{eqn:srrmsr}
\end{equation}
where $\bl E = c\bl Z + d\bl  Z^\top $, with
$\bl Z\sim N_{n\times n}(\bl 0,\bl I)$,
 $c= \sigma \{ (1+\rho)^{1/2} + (1-\rho)^{1/2}\}/2$ and
 $d= \sigma\{(1+\rho)^{1/2} - (1-\rho)^{1/2}\}/2$.
Then $\bl E$ is a mean-zero  Gaussian matrix with 
$\Var{   ( \begin{smallmatrix} e_{i,j} \\ e_{j,i}   \end{smallmatrix} ) } 
 = \sigma^2 (\begin{smallmatrix} 1 & \rho  \\ \rho & 1 \end{smallmatrix}) \equiv \Sigma_e$ and
$\Var{ e_{i,i} }  = \sigma^2 (1 + \rho) $, with the
elements of $\bl E$ being otherwise independent.
Now given $\bs\beta$, $\bl a$ and $\bl b$, construct 
 $\bl E = \bl Y - (  \bl M(\bl X, \bs\beta)  + \bl a\bl 1^\top + \bl 1 \bl b^\top  )$. 
As a function of $\sigma^2$ and $\rho$, the density of
$\bl E$ is proportional to
\[
    (\sigma^2)^{-n^2/2}  (1-\rho^2)^{-{ n \choose 2}/2 } 
    (1+\rho)^{-n/2 }   \times 
    \exp\{  - ( SS_1 + SS_2 )/[2\sigma^2] \} \]
where
$
SS_1 =  \sum_{i<j}(  \begin{smallmatrix} e_{i,j} \\ e_{j,i} \end{smallmatrix} )^
\top ( \begin{smallmatrix} 1 & \rho \\ \rho & 1  \end{smallmatrix} )^{-1} (  
\begin{smallmatrix} e_{i,j} \\ e_{j,i} \end{smallmatrix} )    \}$ and
$SS_2 =  \sum_{i=1}^n  e_{i,i}^2/(1+\rho )$. 
The full conditional distribution of  $1/\sigma^2$ is therefore
$\text{gamma}( [\nu_0 + n^2 ]/2, [ \nu_0 \sigma_0^2 + SS_1 +SS_2  ]/2 )$.
As for $\rho$, 
we do not know of a standard semiconjugate prior distribution. 
However, $\rho$  is just a scalar parameter bounded between
-1 and +1, and so approximate simulation of $\rho$ from its
full conditional distribution (given an arbitrary prior
distribution) could be achieved by computing the unnormalized posterior
density on a grid of values, or by slice sampling, or instead
using a Metropolis-Hastings updating procedure.

To update $\Sigma$ in step 4, 
let $\bl f_i = (\bl a_i, \bl b_i)$ 
and recall that the random
effects model for the $\bl f_i$'s is that
$\bl f_1,\ldots, \bl f_n\sim $ i.i.d.\! $N_2(0,\Sigma)$.
Given a Wishart prior distribution for $\Sigma^{-1}$,
the conditional distribution of $\Sigma^{-1}$ given
$\bl f_1,\ldots, \bl f_n$ is
Wishart$( [ \eta_0\Sigma_0 +\bl F^\top \bl F]^{-1} , \eta_0 + n )$,
where $\bl F$ is the $n\times 2$ matrix with $i$th row equal to $\bl f_i$.

The missing entries of $\bl Y$ may be updated by simulating 
from their full conditional distributions. 
The  full conditional distribution of diagonal entry $y_{i,i}$ is 
$N( m_{i,j} + a_i + b_j, \sigma^2 (1+\rho))$. 
If a dyadic pair of outcomes $(y_{i,j}, y_{j,i})$  is missing, 
then its full conditional distribution is 
bivariate normal
with mean vector 
$(m_{i,j} + a_i + b_j, m_{j,i} + a_j + b_i)$ and 
covariance matrix $\sigma^2 (\begin{smallmatrix} 
       1 & \rho \\ \rho & 1 \end{smallmatrix} ) )$. 
However, if $y_{i,j}$ is observed and $y_{j,i}$ is not, then 
the full conditional distribution of $y_{j,i}$ is normal 
with mean $\rho\times (y_{i,j} - m_{i,j} - a_i -b_j) + 
                m_{j,i} + a_j+ b_i$ and 
variance $\sigma^2(1-\rho^2)$.

Step 1 of the Gibbs sampler requires
 simulation of $\{\bs\beta, \bl a, \bl b\}$
from its
joint  distribution given $\bl Y$, $\Sigma$, $\sigma^2$,  and $\rho$. 
This is challenging because of the dyadic correlation.
However, calculations are simplified by transforming $\bl Y$
so that the dyadic correlation is zero:
Given values of $\sigma^2$ and $\rho$, we may construct
$\tilde {\bl Y} =  \tilde c \bl Y + \tilde d \bl Y^\top$,
where
$\tilde c= \{ (1+\rho)^{-1/2} + (1-\rho)^{-1/2}\}/(2\sigma)$ and
$\tilde d=  \{(1+\rho)^{-1/2} - (1-\rho)^{-1/2}\}/(2\sigma).$
It follows that
\begin{equation} 
\label{eqn:srrmsrdc}
\tilde{\bl Y} \stackrel{d}{=}   \bl M(\tilde {\bl X}, \bs\beta) + \tilde{\bl a}\bl 1^\top + \bl 1 \tilde{\bl b}^\top + \bl Z , 
\end{equation}
where
$\bl Z\sim N_{n\times n}(\bl 0 , \bl I)$,
$\tilde {\bl x}_{i,j} =  \tilde c \bl x_{i,j} + \tilde d \bl x_{j,i}$ ,
$(\tilde a_1,\tilde b_1),\ldots, (\tilde a_n,\tilde b_n)\sim \, \text{i.i.d} \,  N_{2}(\bl 0 , \tilde \Sigma)$ with
 $\tilde \Sigma = \Sigma_e^{-1/2} \Sigma \Sigma_e^{-1/2}$.
Therefore, simulation of $\{ \bs\beta, \bl a , \bl b\}$ from its
conditional distribution given $\bl Y, \Sigma, \sigma^2$ and $\rho$
may be accomplished as follows:
\begin{enumerate}
\item[1.a]  Compute $\tilde {\bl Y}$, $\tilde {\bl X}$
and $\tilde \Sigma = \Sigma_e^{-1/2} \Sigma \Sigma_e^{-1/2}$; 
\item[1.b] Simulate $\{\bs\beta, \tilde{\bl a}, \tilde {\bl b} \}$
from its conditional distribution based on
(\ref{eqn:srrmsrdc});
\item[1.c] Set $(\begin{smallmatrix} a_i \\ b_i \end{smallmatrix} ) = \Sigma_e^{1/2}  (\begin{smallmatrix} \tilde a_i \\ \tilde b_i \end{smallmatrix} ) $
for $i=1,\ldots, n$.
\end{enumerate}

Step 1.b may be implemented by simulating 
$\bs\beta$ conditional on $\{\tilde {\bl Y}, \tilde {\bl X}, \tilde{\Sigma}\}$
and then simulating $\{\tilde{\bl a}, \tilde{\bl b}\}$ conditional on 
$\bs\beta$ and  $\{\tilde {\bl Y}, \tilde {\bl X}, \tilde{\Sigma}\}$. 
We first derive the latter distribution, as it 
facilitates the derivation of the former. 
For notational simplicity,
we drop the tildes on the symbols.

Let $\bl Y = \bl M + \bl a \bl 1^\top  + \bl 1\bl b^\top + \bl Z$
where the elements of
$\bl Z$ are i.i.d.\ standard normal random variables, and 
let $\bl f = (\bl a , \bl b)$ be the concatenation of $\bl a$ and $\bl b$ so 
that 
$\bl f \sim N_{2n}(\bl 0, \Sigma\otimes \bl I)$, 
where ``$\otimes$'' denotes the Kronecker product. 
Vectorizing the formula for $\bl Y$ gives
$ 
\bl y = \bl  m + [ ( \bl 1 \otimes \bl  I ) \  (\bl  I\otimes \bl  1) ]\bl  f
 + \bl z . $
Let $\bl r=\bl y-\bl m$ and $\bl W = [ ( \bl 1 \otimes \bl  I ) \  (\bl  I\otimes \bl  1) ]$. The conditional density of $\bl f$ given
$\bl r$ and $\Sigma$ is given by
\begin{align*}
p(\bl f | \bl r , \Sigma) & \propto 
 \exp(-(\bl r - \bl W \bl f)^\top (\bl r - \bl W \bl f)/2 ) \times 
 \exp(-\bl f^\top ( \Sigma^{-1}\otimes \bl I ) \bl f /2 ) \\
&\propto 
 \exp( -\bl f^\top [ \bl W^\top \bl W + \Sigma^{-1}\otimes \bl I ] \bl f/2 
     + \bl f^\top \bl W^\top \bl r ). 
\end{align*} 
This is the kernel
of a multivariate normal distribution with 
variance $\Var{\bl f|\bl r} = ( \bl W^\top \bl W + \Sigma^{-1} \otimes \bl I )^{-1}$ and expectation $\Exp{\bl f|\bl r} = ( \bl W^\top \bl W + \Sigma^{-1} \otimes \bl I )^{-1}  \bl W^\top \bl r$. 
Some matrix manipulations yield  
$\Var{\bl f|\bl r} = \bl G \otimes \bl  I - \bl H \otimes \bl 1\bl 1^\top$,
where
\begin{itemize}
\item $\bl G = ( \Sigma^{-1} + n \bl I )^{-1} $; 
\item $\bl H = (\Sigma^{-1} +  n \bl 1\bl 1^\top )^{-1} ( \begin{smallmatrix} 0 & 1 \\ 1 & 0 \end{smallmatrix}   )  \bl G$. 
\end{itemize}
Now let $\bl s= \bl W^\top \bl r =
  ( \bl 1^\top \bl R^\top , \bl 1^\top \bl R)$, the
concatenation of the row sums and column sums of $\bl R=\bl Y- \bl M$.
We then have
$\Exp{\bl f |\bl r}  = (\bl G\otimes \bl I )\bl  s - (\bl H \otimes\bl  1\bl 1^\top )\bl  s$. Writing this in terms of the $n\times 2$ matrix $\bl F$
whose vectorization is $\bl f$, 
we have
$\Exp{\bl F | \bl R } =  \bl S \bl G - t \bl 1 \bl  1^\top \bl H$, 
 where $\bl S$ is the $n\times 2$ matrix whose first and second
columns are the row and column sums of $\bl R$, respectively,
and $t = \bl 1^\top \bl R \bl 1$, the sum total of the entries of $\bl R$.
Therefore, to simulate $\bl F$ (and hence $\bl a$ and $\bl b$) from its full conditional
distribution, we set $\bl F$ equal to
\[
 \bl F  = ( \bl S \bl G - t \bl 1 \bl  1^\top \bl H )  + \bl E 
\]
where $\bl E$ is a simulated $n\times 2$ normal  matrix with mean
zero and variance  $\bl G \otimes \bl  I - \bl H \otimes \bl 1\bl 1^\top$.
To simulate this normal matrix, rewrite
$\Var{\bl f|\bl r}$ as
$\Var{\bl f|\bl r} = [\bl G - n\bl H]\otimes \bl I  + n \bl H\otimes [\bl I-\bl 1\bl 1^\top/n]$, and recognize this as the covariance matrix of
\[
  \bl Z_1 ( \bl G - n\bl H)^{1/2} + (\bl I - \bl 1\bl 1^\top/n) \bl Z_2  (\sqrt n \bl H)^{1/2}, \]
where $\bl Z_1$ and $\bl Z_2$ are both $n\times 2$  matrices of
standard normal entries.
To summarize,
to simulate $\bl F$ from its full conditional distribution,
\begin{enumerate}
\item  Simulate two $n\times 2$ matrices $\bl Z_1$ and $\bl Z_2$
 with i.i.d.\ standard normal entries;
\item Compute $\bl E =  \bl Z_1 ( \bl G - n\bl H)^{1/2} + (\bl I - \bl 1\bl 1^\top/n) \bl Z_2  (\sqrt n \bl H)^{1/2}$;
\item Set $\bl F =  ( \bl S \bl G - t \bl 1 \bl  1^\top \bl H )  + \bl E $.
\end{enumerate}

We can use this result to obtain the conditional distribution of 
 $\bs\beta$ given $\bl y$  and $\Sigma$ (but unconditional on $\bl a$, $\bl b$). The density of this distribution  
is proportional to $p(\bl y | \bs\beta,\Sigma ) \pi(\bs\beta)$, 
the product of the SRRM likelihood and the prior density for $\bs\beta$. 
The SRRM likelihood may be obtained using Bayes' rule, 
$p(\bl y| \bs\beta, \Sigma) = p(\bl y|\bs \beta, \bl a, \bl b) p(\bl a, \bl b|\Sigma)/ p(\bl a, \bl b| \bl y, \bs\beta, \Sigma)$. 
The terms on the right side of this equation are easily available:
$p(\bl y|\bs \beta, \bl a, \bl b)$ is  the product of
univariate normal densities corresponding to $y_{i,j} \sim N( \bs \beta^\top \bl x_{i,j} + a_i + b_j, 1)$ independently across ordered pairs.
The terms $p(\bl a, \bl b |\Sigma)$ and
 $p(\bl a, \bl b |\bl y,\bs\beta,\Sigma)$  are the prior
and full conditional distributions of $(\bl a , \bl b)$, the latter
having been obtained in the previous paragraph.
Putting these terms together and simplifying yields the following
form for the uncorrelated SRRM likelihood:
\begin{align} 
\label{eqn:srmlik}
p(\bl y | \bs\beta, \Sigma )  &= 
(2\pi )^{-n^2/2}   
  | \bl I+n  \Sigma|^{-(n-1)/2} | \bl I + n \Sigma \bl 1\bl 1^\top |^{-1/2}  \times   \nonumber \\
&  \exp\{ -({\bl r}^\top {\bl r } + t^2 \bl 1^\top \bl H\bl 1   - \text{tr}(\bl S^\top \bl S \bl G)  )/2 \}. \nonumber 
\end{align}
This is quadratic in the $r_{i,j}$'s, and hence also quadratic in
$\bs\beta$.
Some algebra gives
\[
 p(\bl y | \bs\beta, \Sigma )   \propto 
  \exp\{ - \bs\beta^\top (\bl Q_1+ \bl Q_2 + \bl Q_3)  \bs\beta /2 + \bs\beta^\top (\bs \ell_1 + \bs\ell_2 +\bs\ell_3)\}, 
\]
where
$\bl Q_1 = {\bl X}^\top {\bl X}$ and
      $\bs \ell_1 = {\bl X}^\top {\bl y}$,
   with ${\bl X}$ being the $n^2 \times p$ matrix of the
    ${\bl x}_{i,j}$'s ;
$\bl Q_2 = n^4 h \bar {{\bl x}} \bar{{\bl x}}^\top$ and
      $\bs\ell_2 = n^4 h \bar {{\bl x}} \bar {y} $
     with $h=\bl 1^\top \bl H \bl 1$,
     $\bar{ {\bl x}}$ being the average of the ${\bl x}_{i,j}$'s
     and    ${\bar{ y}}$ being the average of the $ y_{i,j}$'s,
and
\begin{align*}
\bl Q_3 &=  - n^2( g_{11} \bar {{ \bl X}}_r^\top \bar {{ \bl X}}_r  + g_{12}(  \bar {{ \bl X}}_r^\top \bar {{ \bl X}}_c  +  \bar {{ \bl X}}_c^\top \bar {{ \bl X}}_r   ) +
               g_{22}  \bar {{ \bl X}}_c^\top \bar {{ \bl X}}_c  ) \\ 
\bs\ell_3 &=  - n^2( g_{11}  \bar {{ \bl X}}_r^\top  {\bar { {\bl y}}}_r      + g_{12} ( {{ \bl X}}_r^\top  {\bar { {\bl y}}}_c  + \bar {{ \bl X}}_c^\top  {\bar { {\bl y}}}_r   )    + g_{22}  \bar {{ \bl X}}_c^\top  {\bar { {\bl y}}}_c       ), 
\end{align*}
where
${\bar {{\bl y}}}_r$
is the $n\times 1$ vector of row means of ${\bl Y}$, 
$\bar{{\bl  X}}_r$ is the  $n\times p$ matrix
whose $i$th row is the average of ${\bl x}_{i,j}$ over
$j=1,\ldots, n$, and
${\bar {{\bl y}}}_c$ and $\bar {{ \bl X}}_c$
are  analogously defined as column means.
Now the prior density for $\bs\beta$ is proportional to 
$ \exp\{  - \bs \beta^\top \bl Q_0 \bs\beta /2  +
   \bs\beta^\top \bl Q_0 \bs\beta_0\}$, and so the conditional
density is given by
\[ p(\bs\beta | \bl y, \Sigma ) \propto  p(\bl y | \bs\beta,\Sigma) \times  
   \pi(\bs\beta) 
 \propto 
  \exp\{ -\bs\beta^\top ( \bl Q_0 + \bl Q) \bs\beta/2
    + \bs\beta^{\top} ( \bl Q_0 \bs\beta_0 +  \bs \ell  ) \} 
\]
where $\bl Q = \bl Q_1+\bl Q_2 + \bl Q_3$ and $\bs \ell = 
 \bs \ell_1+\bs \ell_2+\bs \ell_3$. This is a multivariate normal
density, with variance $(\bl Q_0+\bl Q  )^{-1}$ and
mean $(\bl Q_0 + \bl Q )^{-1} (   \bl Q_0 \bs\beta_0 + \bs\ell )$.

\subsection{Gibbs sampling for the AME}
Now suppose that $\bl Y$ follows a Gaussian AME model,
so that
$\bl Y = \bl M(\bl X, \bs\beta) +  \bl U \bl V^\top + 
\bl a \bl 1^\top + \bl 1 \bl b^\top +  \bl E$ where
the distribution of $\{\bl a, \bl b, \bl E\}$ follows the
social relations covariance model with parameters $\{\Sigma, 
\sigma^2, \rho\}$. Let $(\bl u_i,\bl v_i)\sim N_{2r}(\bl 0 , \Psi)$ 
independently across nodes, and let
$\Psi^{-1} \sim$ Wishart$(\Psi_0^{-1}/\kappa_0,\kappa_0)$ {\it a priori}. 
The joint posterior distribution of 
the unknown parameters may be approximated by 
a Gibbs sampler that iterates the 
following steps: 
\begin{enumerate}
\item  Update $(\bs\beta,\bl a, \bl b, \sigma^2,\rho, \Sigma)$ 
and the missing values of $\bl Y$ using the algorithm described in 
Section 6.1, but with $\bl Y$ replaced by $\bl Y - \bl U \bl V^\top$;  
\item  Simulate $\Psi^{-1} \sim \text{Wishart}(  ( \Psi_0\kappa_0 + [ \bl U\, \bl V ]^\top [ \bl U \, \bl V]  )^{-1} ,  \kappa_0+n)  $,  where $[\bl U \, \bl V]$ is the $n\times 2r$ matrix
equal to the column-wise concatenation of $\bl U$ and $\bl V$; 
\item  For each $k=1,\ldots, r$, simulate the $r$th columns of 
$\bl U$ and $\bl V$ from their full conditional distributions; 
\end{enumerate}
To perform step 3, first consider
the full conditional distribution of $\bl u_1$, the first column of
$\bl U$. Let
$\bl R = \bl Y -  ( \bl M(\bl X, \bs\beta)  +  \sum_{k=2}^r \bl u_k \bl v_k^\top +    \bl a \bl 1^\top   +\bl 1 \bl b^\top )$. Then we have
$ \bl R  = \bl u_1 \bl v_1^\top  + \bl  E $. Decorrelating 
gives $\tilde {\bl R} = \tilde c \bl R + \tilde d \bl R = \tilde c \bl u_1 \bl v_1^\top + 
 \tilde d \bl v_1 \bl u_1^\top + \bl Z $, 
and
vectorizing gives
$\tilde{\bl r} = [ \tilde c (\bl v_1 \otimes \bl I) + \tilde d ( \bl I \otimes \bl v_1)  ] \bl u_1 + \bl z $. 
Given $\bl v_1$,  this is a linear regression model 
with outcome vector $\tilde{\bl r}$, 
 design matrix $\bl W = [ \tilde c (\bl v_1 \otimes \bl I) + \tilde d ( \bl I \otimes \bl v_1)  ]$, regression parameters $\bl u_1$, and i.i.d.\ standard normal
errors.  Let $\bs \mu_{u|v}$ and $\Sigma_{u|v}$ be the conditional mean 
and variance of $\bl u_1$ given $\bl v_1$. 
Then the conditional distribution of $\bl u_1$ given
$\bl v_1$ and $\tilde{\bl R}$ is normal with mean and variance given by 
\begin{align*}
\Var{\bl u_1| \tilde{\bl R}, \bl v_1}  & = ( \Sigma_{u|v}^{-1} + \bl W^\top \bl W )^{-1} \\
\Exp{\bl u_1| \tilde{\bl R}, \bl v_1} & = 
( \Sigma_{u|v}^{-1} + \bl W^\top \bl W )^{-1}
   ( \Sigma_{u|v}^{-1} \bs \mu_{u|v}  + \bl W^\top \tilde{\bl r}) . 
\end{align*}
Some calculations show that 
$\bl W^\top \bl W = (\tilde c^2 + \tilde d^2) ||\bl v_1||^2 \bl I + 2 \tilde c \tilde d 
  \bl v_1 \bl v_1^\top$ and 
$\bl W^\top \tilde{\bl r}= ( \tilde c \tilde {\bl R} + \tilde d \tilde {\bl R}^\top ) \bl v_1$. 
The full conditional distribution of $\bl v_1$, and the other columns
of $\bl U$ and $\bl V$, may be obtained similarly.

\subsection{Gibbs sampling for transformation models} 
A transformation model assumes that the  sociomatrix $\bl S$ 
is a function of a latent sociomatrix 
$\bl Y$ that follows a Gaussian AME model 
with parameters $\bs\theta = (\bs\beta, \bl a, \bl b, \bl U, \bl V,  \rho, \Sigma, \Psi)$. 
This collection of parameters does not include $\sigma^2$, because for probit models in general  and for the other transformation models 
described in this article, the overall scale of the $y_{i,j}$'s is 
not identifiable, and so we fix $\sigma^2=1$.  
For the transformation models discussed in Section 4, observation of 
$\bl S$ implies that $\bl Y\in C(\bl S)$. 
Given starting values of $\bl Y$ and $\bs\theta$, a Gibbs sampler 
for approximating the joint posterior distribution of $\bl Y$  and $\bs\theta$
conditional on $\bl S$ proceeds by iterating the following steps: 
\begin{enumerate}
\item Update $\bs\theta$ conditional on $\bl Y$ with
the algorithm described in Section 6.2; 
\item Update $\bl Y$ conditional on $\bs\theta$ and $\bl Y\in C(\bl S)$. 
\end{enumerate}

To perform step 2 of this algorithm, 
first consider the simple probit transformation model where 
the observed outcome $s_{i,j}$ is the binary indicator that 
the latent Gaussian variable $y_{i,j}$ is greater than zero. 
Let $\mu_{i,j} = \bs\beta^\top \bl x_{i,j} + \bl u_i^\top \bl v_j + a_i + b_j$.
Then unconditional on $\bl S$ but given the other parameters, we have that
\begin{align*}
\begin{pmatrix} y_{i,j} \\ y_{j,i}  \end{pmatrix}  \sim N_2 \left( 
\begin{pmatrix} \mu_{i,j} \\ \mu_{j,i}  \end{pmatrix} , 
  \begin{pmatrix} 1  &  \rho \\ \rho  & 1  \end{pmatrix}  \right )
\end{align*}
independently 
across dyads, and that $y_{i,i} \sim N( \mu_{i,i} ,1+\rho)$
independently across diagonal entries. Since the diagonal entries of
$\bl S$ are undefined and the diagonal entries of $\bl Y$ are uncorrelated
with the off-diagonal entries, each $y_{i,i}$ value may be updated
from its $N(\mu_{i,j} ,1+\rho)$ distribution.
The off-diagonal entries may be updated in two steps: first
updating the elements of $\bl Y$ below the diagonal, and then
updating those above. To do so, note that $y_{i,j}| y_{j,i} \sim 
N(\mu_{i,j} + \rho \times (y_{j,i}-\mu_{j,i}),1-\rho^2)$.
Now in the case of a probit AME model where
$s_{i,j}$ is the indicator that $y_{i,j}$ is greater than zero,
the full conditional distribution of $y_{i,j}$ is
$N(\mu_{i,j} + \rho (y_{j,i}-\mu_{j,i}),1-\rho^2)$ 
but
constrained to be above zero if $y_{i,j}=1$ and below zero
otherwise. The full conditional distributions under other
types of transformation models are also constrained normal distributions,
where the constraint depends on the type of transformation.
Univariate constrained normal distributions may be easily simulated from 
using the inverse-CDF method.

\section{Discussion}
The AME framework is a modular
approach for network data analysis
based on three statistical models: 
the social relations covariance model, 
low-rank matrix representations via multiplicative 
factors, and Gaussian transformation models. 
Separately, each of these should be familiar 
to an applied statistician or data analyst: 
The first is a type of linear random effects model, 
the second is analogous to a model-based singular value 
decomposition, and the third forms the basis 
of many binary and ordinal regression models. 
Together, they provide a flexible model-based framework 
for inference that accounts for 
many statistical dependencies often found in network data, 
and accommodates a variety of types of dyadic and nodal variables. 
Current and future work in this area includes 
generalizing this framework to analyze datasets 
from more modern network studies that include 
multiple sociomatrices on one or more nodesets, 
such as comparison studies across multiple 
populations, multiple time points, multiple 
dyadic variables, or combinations of these. 
Some steps in this direction have been taken 
by representing a set of sociomatrices as a 
tensor \citep{hoff_2011a,hoff_2016}, but these methods 
are not yet general enough to encompass the wide variety 
of multivariate, multilevel and longitudinal network 
datasets that are becoming more prevalent. What is needed 
is a broad framework like that which is provided for 
generalized linear mixed models 
by 
the {\tt nlme} or {\tt lme4} software 
\citep{pinheiro_bates_2000,walker_et_al_2015}, 
whereby a data analyst may separately select
the type of data being analyzed (continuous,
binary, count, etc.) and build a complicated 
model of dependence relationships between subsets of the data. 
One challenge to developing such a framework 
for network data is  computational - 
the Gibbs samplers described in this article  
and implemented in the {\sf R} package {\tt amen} 
become cumbersome when the number of nodes 
is above a few thousand, and other integral approximation 
methods (such as Laplace approximations) for AME transformation models 
are infeasible because of the complicated 
dependence induced by the SRM. 
Fast, stable parameter estimation for 
large network datasets may require abandoning use 
of the full likelihood, and instead use 
composite likelihood 
estimation \citep{lindsay_1988}
or modern method-of-moments approaches \citep{perry_2017}.

\section*{Acknowledgments}
This research was partially supported by NSF grant DMS-1505136.

\bibliography{netreview}

\end{document}